\documentclass[journal]{IEEEtran}
\ifCLASSINFOpdf
\else
   \usepackage[dvips]{graphicx}
\fi
\usepackage{url}

\usepackage{graphicx}

\usepackage{amsmath,amssymb,threeparttable,placeins,makecell,stfloats,cite}
\usepackage[usenames]{color}
\usepackage{amsfonts}
\usepackage{latexsym}
\usepackage{lipsum}
\usepackage{floatrow}
\usepackage{cuted}
\newtheorem{theorem}{{{\textit{Theorem}}}}

\newtheorem{lemma}{{{\textit{Lemma}}}}
\newtheorem{corollary}{{{{\textit{Corollary}}}}}

\newtheorem{definition}{{{\textit{Definition}}}}

\newtheorem{remark}{{{\textit{Remark}}}}

\newtheorem{example}{{{\textit{Example}}}}

\begin{document}
\title{Direct Construction of Optimal Z-Complementary Code Sets for all Possible Even Length by Using Pseudo-Boolean Functions }
\author{Gobinda~Ghosh,~
        Sudhan~Majhi,~
        Palash~Sarkar,~and~Ashish~Kumar~Upadhyay}
\IEEEpeerreviewmaketitle
\maketitle
\begin{abstract}
Z-complementary code set (ZCCS) are well known
to be used in multicarrier code-division multiple access (MCCDMA) system to provide a interference free environment. Based
on the existing literature, the direct construction of optimal ZCCSs
are limited to its length. In this paper, we are interested in constructing optimal ZCCSs of all possible even  lengths using Pseudo-Boolean functions. The maximum column sequence peak-to-man envelop power ratio (PMEPR) of the proposed ZCCSs is upper-bounded by two, which may give an extra benefit in managing PMEPR in an ZCCS based MC-CDMA system, as well as the ability to handle a large number of users.
\end{abstract}
\section{Introduction}
\label{sec:intro}
\IEEEPARstart{M}{ulticarrier}  code-division multiple access (MC-CDMA) is a multiple access scheme used in orthogonal frequency division multiplexing (OFDM)-based telecommunication systems, allowing the system to support multiple users at the same time over the same frequency band. When the number of users in a channel increases, it is found that the performance of MC-CDMA degrades as a result of multi-user interference (MUI) and multipath interference (MPI).
The Complete-complementary code (CCC) \cite{liu2014new} has perfect cross-and auto-correlation characteristics, which allows for simultaneous interference-free transmission in the multi-carrier-digital mobile (MC-CDMA) system.
A major disadvantage of CCC is that the number of supported users is limited by the number of row sequences in each complementary matrix. The set size of the ZCCS system is much bigger than that of the CCC system \cite{sarkar2018optimal}, which enables for a considerably greater number of users to be supported by a ZCCS-based MC-CDMA system, as opposed to a CCC-based MC-CDMA system, where the number of subcarriers is equal to the number of users.\\
In recent literature, research on generalized Boolean functions (GBFs) based constructions of complementary sequences has received great attention from the sequence design community,
\cite{sarkar2018optimal} \cite{wu2018optimal}, \cite{rathinakumar2008complete}, \cite{davis1999peak}, \cite{paterson2000generalized}, \cite{sarkar2020direct},\cite{sarkar2020construction}. The GBFs based construction of CCCs were extended to optimal ZCCSs in \cite{sarkar2018optimal} and \cite{wu2018optimal}. However, GBFs based construction of optimal ZCCSs has a limitation on the sequence lengths which is in the form of power-of-two \cite{sarkar2018optimal},\cite{wu2018optimal},\cite{sarkar2020direct} and \cite{wu2020z}. By extending the idea of GBFs to PBFs, 
recenly, a direct construction of optimal ZCCSs has been introduced in \cite{sarkar2021pseudo} which is able to provide non-power-of-two length sequences but limited to the form  $p2^m$, where $p$ is
a prime number and $m$ is a positive integer.  Another direct construction
like GBFs based constructions, PBFs based constructions are also known as direct constructions in the literature. Direct constructions are feasible for rapid hardware generation \cite{sarkar2018optimal} of sequences.
Besides direct constructions, many indirect constructions can be found in \cite{das2020new}, \cite{b}, \cite{c}, \cite{a} and \cite{d} which are dependent on some kernel at its initial stages.
The limitation on the lengths of optimal ZCCS through direct constructions in the existing literature motivates us in searching of PBFs to provide more flexiblity on the lengths. In search of new ZCCS, in this paper,  we
have proposed a direct construction of optimal ZCCS for all
possible even length using PBFs. 
We also have showed that, the proposed construction is  able to maintain a minimum coloumn sequence PMPER 2, unlike the existing direct construction of optimal ZCCSs of non-power-of-two lengths. The PBF  reported in \cite{sarkar2021pseudo}  appears as a special case of proposed construction.\\
The rest of this work is organised in the following way. In Section II, we will go over a few definitions. Section III offers a comprehensive description of the ZCCS's construction. Section IV of this article ends the study by comparing our findings to those of previous researchers.
\section{Preliminary}
This section presents a few basic definitions and lemmas for use in the proposed construction.
Let 
$\textbf{x}_1=[x_{1,0},x_{1,1},\hdots, x_{1,N-1}]$ and $\textbf{x}_2=[x_{2,0},x_{2,1},\hdots ,x_{2,N-1}]$ be a pair of  sequences whose components are  complex numbers. 
Let $\tau$ be an integer, we define \cite{sarkar2018optimal}
\begin{equation}\label{equ:cross}
\Theta(\textbf{x}_1, \textbf{x}_2)({\tau})=\begin{cases}
\sum_{i=0}^{N-1-\tau}x_{1,i+\tau}x^{*}_{2,i}, & 0 \leq \tau < N, \\
\sum_{i=0}^{N+\tau -1}x_{1,i}x^{*}_{2,i-\tau}, & -N< \tau < 0,  \\
0, & \text{otherwise},
\end{cases}
\end{equation}
and when $\textbf{x}_1=\textbf{x}_2$, $\Theta(\textbf{x}_1,\textbf{x}_2)(\tau)=\mathcal{A}_{{\textbf{x}_{1}}}(\tau)$. This functions $\Theta$ and $\mathcal{A}$ are known as aperiodic cross-correlation function (ACCF) of $\textbf{x}_{1}$ and $\textbf{x}_{2}$ and aperiodic auto-correlation function (AACF) of $\textbf{x}_{1}$ respectively and $*$ denotes the complex conjugate.\\
Let $\mathbf{B}=\{B^{0},B^{1},\hdots,B^{M-1}\}$ be a collection of $M$ matrices each of dimensions $K\times N$, i.e,
    $B^{\delta}=
\big[\mathbf{b}_{0}^{\delta},
\mathbf{b}_{1}^{\delta},
\hdots,
\mathbf{b}_{K-1}^{\delta}\big]^{T}_{K\times N}$,
where the notation T is used to denote the
transpose of a matrix and
 each $\mathbf{b}_{i}^{\delta}$ is a complex-valued sequences of length $N$ i.e, $\mathbf{b}_{i}^{\delta}=(\mathbf{b}_{i,0}^{\delta},\mathbf{b}_{i,1}^{\delta},\hdots, \mathbf{b}_{i,N-1}^{\delta})$.
Suppose $B^{\delta_{1}},B^{\delta_{2}} \in \mathbf{B}$, where $0\leq \delta_{1},\delta_{2}\leq M-1$,
we define the ACCF between $B^{\delta_1}$ and $B^{\delta_2}$ as,
$
    \Theta(B^{\delta_1},B^{\delta_2})(\tau)=\displaystyle\sum_{i=0}^{K-1}\Theta(\mathbf{b}_{i}^{\delta_{1}},\mathbf{b}_{i}^{\delta_{2}})(\tau).
$
When the following equation holds, we refer to the set $\mathbf{B}$ as ZCCS.

\begin{definition}(\cite{sarkar2018optimal})
Code set $\mathbf{B}$ is called a ZCCS if
\begin{eqnarray}
\Theta(B^{\delta_1},B^{\delta_2})(\tau)
=\begin{cases}
KN, & \tau=0, \delta_1=\delta_2,\\
0, & 0<|\tau|<Z, \delta_1=\delta_2,\\
0, & |\tau|< Z, \delta_1\neq \delta_2,
\end{cases}
\end{eqnarray}
\end{definition}
where $Z$ denotes ZCZ width. With the parameter $K,N,M$ and $Z$, we denote the set of matrices $\mathbf{B}$ as a $(M,Z)-ZCCS_{K}^{N}$, which is called optimal if $M=K\lfloor \frac{N}{Z}\rfloor$ and non-optimal if 
$M<K\lfloor \frac{N}{Z}\rfloor$ \cite{liu2011correlation}.  When $K=M$ and $Z=N$, we denote $\mathbf{B}$ by $(K,K,N)$-CCC.
\vspace{0.1cm}
\begin{lemma}(Constrction of CCC\cite{rathinakumar2008complete})\label{lemma3}\\
Let $g:\mathbb{Z}_2^m\rightarrow \mathbb{Z}_q$ be a second-order GBF and $\tilde{g}$ be the reversal of $g$ i.e, $\tilde{g}(y_{0},y_{1},\hdots, y_{m-1})\!=\!g(1\!-\!y_{0},1\!-\!y_{1},\hdots, 1\!-\!y_{m-1})$. Assume that  the graph   $G(g)$ contains vertices denoted as $y_{\beta_{0}},y_{\beta_{1}},\hdots, y_{\beta_{n-1}}$ such that, after executing a deletion operation on those vertices, the resultant graph is reduced to a path. We define the weight of each edge by $\frac{q}{2}$. Let the binary representation of the integer ${r}$ is $\mathbf{r}=(r_0,r_1,\hdots,r_{n-1})$. Define the $G_r$ and $\bar{G}_r$ to be
\begin{equation}
\begin{split}
    &\Big\{ {g}\!+\!
 \frac{q}{2}\Big((\mathbf{v}+\mathbf{r})\cdot {\mathbf{y}}
 \!+\!{v_{n}}y_{\gamma}\Big): \mathbf{v}\in \{0,1\}^n,v_{n}\in\{0,1\}\Big\},\\
 &\Big\{ \tilde{g}\!+\!
 \frac{q}{2}\Big((\mathbf{v}+\mathbf{r})\cdot \bar{\mathbf{y}}
 \!+\!\bar{v}_{n}y_{\gamma}\Big): \mathbf{v}\in \{0,1\}^n,v_{n}\in\{0,1\}\Big\},
\end{split}
 \end{equation}
 respectivly. Denote  $(\cdot)\cdot(\cdot)$ as the  dot product of two real-valued vectors  $(\cdot)$ and $(\cdot)$, $\gamma$  specifies the label for either of the path's end vertices. $\mathbf{y}=(y_{\beta_{0}},y_{\beta_{1}},\hdots, y_{\beta_{n-1}})$, 
 $\bar{\mathbf{y}}=(1-y_{\beta_{0}},1-y_{\beta_{1}},\hdots,1-y_{\beta_{n-1}})$, and $\mathbf{v}=(v_0,v_1,\hdots,v_{n-1})$.
 Then
  $\{\Psi(G_r),\Psi^*(\bar{G}_{r}):0\leq r<2^n\}$ forms
$(2^{n+1},2^{n+1},2^m)$-CCC, where $\Psi^*(\cdot)$ denotes the complex conjugate of $\Psi(\cdot)$.
\end{lemma}
\subsection{Pseudo-Boolean Functions (PBFs)}
A degree $i$ monomial is a product of $i$ distinct varibles among $y_0,y_1,\hdots,y_{m-1}$. PBFs are functions $\mathbf{F}:\{0,1\}^{m}\rightarrow\mathbb{R}$,  that are represented as a linear combination of monomials among $\{y_0,y_1,\hdots, y_{m-1}\}$ where, $y_i$'s are Boolean variable and coeffecients are drawn from $\mathbb{R}$. The highest degree of the monomials are called the degree of $\mathbf{F}$. As an example $\frac{4}{3}y_2y_1+y_0$ is a 2nd order PBF of three variables $y_0,y_1$ and $y_2$.
It is clear that, when this coeffecients are drawn from $\mathbb{Z}_{q}$ and the range of the fuction $\mathbf{F}$ changed to $Z_{q}$ the function $\mathbf{F}$ becomes a generalized Boolean function (GBF)\cite{sarkar2021pseudo}.
Let $l$ be a positive integer and $p_1,p_2,\hdots, p_l$ be prime numbers  and $\mathbf{c}=({c}_1,{c}_2,\hdots, {c}_l)$, where ${c}_i \in \{0,1,\hdots, p_i\!-\!1\}$. Let $g$ be a 2nd order boolean function  of $m$ variables and let $\mathbf{{Y}}=(y_{0},\hdots,y_{m+\sum_{i=1}^{l}s_i-1})$. We define the following PBFs with the help of $g$ as\\
\begin{equation}
    \begin{split}
       &M^{\mathbf{c}}(\mathbf{{Y}})\!=\!g(y_{0},\hdots,y_{m-1})+\displaystyle\sum_{i=1}^{l}\frac{{c}_{i}q}{p_{i}}\displaystyle\sum_{k=0}^{s_{i}-1}2^{k}y_{m+{\sum_{j=0}^{i-1}s_{j}+k}}\\
        &N^{\mathbf{c}}(\mathbf{{Y}})\!=\!\tilde{g}(y_{0},\hdots,y_{m-1})+\displaystyle\sum_{i=1}^{l}\frac{{c}_{i}q}{p_{i}}\displaystyle\sum_{k=0}^{s_{i}-1}2^{k}y_{m+{\sum_{j=0}^{i-1}s_{j}+k}} 
    \end{split}
\end{equation}
where $s_{i}\in \mathbb{Z}^{+}$ which denotes the set of all positive integer, $s_{0}=0$ and each $y_{i}$'s are Boolean variable. From (4), it is clear 
that both $M^{\mathbf{c}}$ and $N^{\mathbf{c}}$ are PBFs of variables $m+\sum_{i=1}^{l}s_i$. We chose $s_i$ in such a way that $p_i\leq 2^{s_i}$ $\forall i\in\{1,2,\hdots,l\}$.
Let  $\!h\!:\!\{0,1\}^{n+1}\rightarrow \mathbb{Z}_{q}$ be a  function  and $\mathbf{r}=(r_0,r_1,\hdots, r_{n-1})$ be the binary representation of the integer $r$ where, $0\leq r<2^n$, and $\mathbf{c}=({c}_1,{c}_2,\ldots, {c}_l)$.
We define the sets as,
\begin{equation}
     \begin{split}
       &\Omega_r^\mathbf{c}\!\!=\!\!\Big\{{M}^\mathbf{c}(\mathbf{Y})+h(\mathbf{v'})\!+\!
 \frac{q}{2}\Big((\mathbf{v}+\mathbf{r})\cdot {\mathbf{y}}
 \!+\!{v_{n}}y_{\gamma} \Big)\\
 &~~~~~~~~~~~~~~~~~~~~~~~~~~~~~~~~~~~:\mathbf{v}\in \{0,1\}^n,v_{n}\in\{0,1\}\Big\},\\
 &\Lambda_r^\mathbf{c}\!\!=\!\!\Big\{{N}^\mathbf{c}(\mathbf{Y})+h(\mathbf{v'})\!+\!
 \frac{q}{2}\Big((\mathbf{v}+\mathbf{r})\cdot {\mathbf{\bar{y}}}
 \!+\!{\bar{v}_{n}}y_{\gamma} \Big)\\
 &~~~~~~~~~~~~~~~~~~~~~~~~~~~~~~~~~~~:\mathbf{v}\in \{0,1\}^n,{v}_{n}\in\{0,1\}\Big\},
     \end{split}
 \end{equation}  
 where, $\mathbf{v'}=(\mathbf{v},v_{n})=(v_{0},v_{1},\hdots,v_{n})$.
Let\! us assume that $g$ be the GBF as defined in \textit{Lemma} \textit{1}. Let $g^{\mathbf{v'},\mathbf{r},h}\!\!=\!\!g+h(\mathbf{v'})+\frac{q}{2}((\mathbf{v}+\mathbf{r})\cdot \mathbf{y}
 \!+v_{n}y_{\gamma})$ and $
 s^{\mathbf{v'},\mathbf{r},h}\!\!=\!\!\tilde{g}+h(\mathbf{v'})+\frac{q}{2}((\mathbf{v}+\mathbf{r})\cdot \bar{\mathbf{y}}
 \!+\bar{v}_{n}y_{\gamma})$ . We also assume $M^{\mathbf{v'},\mathbf{r},\mathbf{c}}=M^\mathbf{c}(\mathbf{Y})\!+\!\frac{q}{2}((\mathbf{v}+\mathbf{r})\cdot \mathbf{y})
 \!+\!v_{n}y_{\gamma}$  and $N^{\mathbf{v'},\mathbf{r},\mathbf{c}}=N^\mathbf{c}(\mathbf{Y})\!+\!
 \frac{q}{2}((\mathbf{v}+\mathbf{r})\cdot \bar{\mathbf{y}}
 \!+\!\bar{v}_{n}y_{\gamma})$. As per our assumption, for any choice of $\mathbf{v'}\in \{0,1\}^{n+1}$ and $\mathbf{r}\in \{0,1\}^n$,
 the functions $g^{\mathbf{v'},\mathbf{r},h}$ and $s^{\mathbf{v'},\mathbf{r},h}$ are $\mathbb{Z}_q$-valued GBFs of $m$ variables and 
 $M^{\mathbf{v'},\mathbf{r},\mathbf{c}}$ and $N^{\mathbf{v'},\mathbf{r},\mathbf{c}}$ are PBFs of $m+\sum_{i=0}^{l}s_i$ variables. Let  $M^{\mathbf{v'},\mathbf{r},\mathbf{c},h}(\mathbf{Y})=M^{\mathbf{v'},\mathbf{r},\mathbf{c}}+h(\mathbf{v'})$ and $N^{\mathbf{v'},\mathbf{r},\mathbf{c},h}(\mathbf{Y})=N^{\mathbf{v'},\mathbf{r},\mathbf{c}}+h(\mathbf{v'})$.
We define $\Psi(M^{\mathbf{v'},\mathbf{r},\mathbf{c},h})$, the complex-valued sequence  as  $ \Psi(M^{\mathbf{v'},\mathbf{r},\mathbf{c},h})=
 (\omega^{M^{\mathbf{v'},\mathbf{r},\mathbf{c},h}_0}_q, \omega^{M^{\mathbf{v'},\mathbf{r},\mathbf{c},h}_1}_q,\hdots,\omega_q^{M^{\mathbf{v'},\mathbf{r},\mathbf{c},h}_{2^{m+\sum_{i=0}^{l}s_i}-1}})$,
 where, $M^{\mathbf{v'},\mathbf{r},\mathbf{c},h}_{r'}\!\!\!=\!\!\!M^{\mathbf{v'},\mathbf{r},\mathbf{c},h}(r_0,r_1,\hdots,r_{m+\sum_{i=0}^{l}s_i-1})$, $0\!\leq\! r'\!<\!2^{m+\sum_{i=0}^{l}s_i}$
and  the binary representation of the integer $r'$ is  $(r_0,r_1,\hdots,r_{m+\sum_{i=0}^{l}s_i-1})$.  The $r'$-th component of $\Psi(M^{\mathbf{v'},\mathbf{r},\mathbf{c},h})$ is given by
\begin{equation}\label{wvvd}
 \begin{split}
&w_{q}^{M_{r'}^{\mathbf{v}',\mathbf{r},\mathbf{c},h}}=\omega_q^{{g}^{\mathbf{v'},\mathbf{r},h}(r_0,r_1,\hdots, r_{m-1})}\omega_{p_1}^{{c}_1\sum_{i=0}^{s_{1}-1}2^{i}r_{m+i}}\\&\omega_{p_2}^{{c}_2\sum_{i=0}^{s_{2}-1}2^{i}r_{m+s_{1}+i}}\ldots\omega_{p_l}^{{c}_l\sum_{i=0}^{s_{l}-1}2^{i}r_{m+\sum_{j=1}^{l-1}s_{j}+i}}\\&=\omega_q^{{g}_{j}^{\mathbf{v'},\mathbf{r},h}}\omega_{p_1}^{{c}_1(i_{1})}\omega_{p_2}^{{c}_2(i_{2})}\ldots\omega_{p_l}^{{c}_l(i_{l})},\\
 \end{split}
\end{equation}
where,  $(r_{0},r_{1},\hdots,r_{m-1})$\! is the binary representation of the integer $j$.
Since $M^{\mathbf{v'},\mathbf{r},\mathbf{c}}$ is a $m\!+\!\sum_{i=1}^{l}s_{i}$ variable PBF therefore  the length of $\Psi(M^{\mathbf{v'},\mathbf{r},\mathbf{c},h})$  is $2^{s_1}2^{s_2}\ldots2^{s_l}2^{m}$. Any element of $\Psi(M^{\mathbf{v'},\mathbf{r},\mathbf{c},h})$ is of the form $w_{q}^{g_{j}^{\mathbf{v'},\mathbf{r},h}}w_{p_1}^{{c}_{1}(i_{1})}w_{p_2}^{{c}_{2}(i_{2})}\hdots
w_{p_l}^{{c}_{l}(i_{l})}$,\! where $0\leq i_{k}\!\leq\!2^{s_{k}\!}-\!1$, $0\leq j\leq 2^{m}-1$ and $0\leq k\leq l$. 
\begin{lemma}({\cite{vaidyanathan2014ramanujan}} )\label{L3} Let $t$ and $t'$ be two non-negative integers, where $0\leq t\neq t'<p_{i}$, $p_{i}$ is a prime number as defined in section-II. 
 Then 
 $\displaystyle\sum_{j=0}^{p_{i}-1}\omega_{p_{i}}^{(t-t')j}=0.$
\end{lemma}
Let $S\!=\!\{ M^{\mathbf{v_{1}'},\mathbf{r},\mathbf{c},h},M^{\mathbf{v_{2}'},\mathbf{r},\mathbf{c},h},\hdots, M^{\mathbf{v_{2^{n+1}}'},\mathbf{r},\mathbf{c},h}\}$ where, $\mathbf{v}'_{k}\in \{0,1\}^{n+1}$ and $k\in\{1,2,\hdots,2^{n+1}\}$. We define
\begin{equation}
    \Psi(S)=[\Psi( M^{\mathbf{v_{1}'},\mathbf{r},\mathbf{c},h}),\Psi( M^{\mathbf{v_{2}'},\mathbf{r},\mathbf{c},h}),\hdots,\Psi( M^{\mathbf{v_{2^{n+1}}'},\mathbf{r},\mathbf{c},h})]^{T}.
\end{equation}
Now we truncate the sequence $\Psi(M^{\mathbf{v'},\mathbf{r},\mathbf{c},h})$  by reomoving  all the elements of the form  $w_{q}^{g_{j}^{\mathbf{v'},\mathbf{r},h}}\!w_{p_1}^{{c}_{1}(i_{1})}\!w_{p_2}^{{c}_{2}(i_{2})}\!\hdots\!
w_{p_l}^{{c}_{l}(i_{l})}$  from 
\newpage
~
 \begin{strip}
 \begin{equation}
    \begin{split}
        &\Psi^{i}_{Trun}(M^{\mathbf{v'},\mathbf{r},\mathbf{c},h})=\underbrace{(w_{p_1}^{{c}_{1}(i_{1})}w_{p_2}^{c_{2}(i_{2})}\hdots
w_{p_l}^{c_{l}(i_{l})})w_{q}^{g_{0}^{\mathbf{v},\mathbf{r},d,h}},\hdots, (w_{p_1}^{c_{1}(i_{1})}w_{p_2}^{c_{2}(i_{2})}\hdots w_{p_l}^{c_{l}(i_{l})})w_{q}^{g_{2^{m}-1}^{\mathbf{v},\mathbf{r},d,h}}}\\
 &\Psi^{i}_{Trun}(N^{\mathbf{v'},\mathbf{r},\mathbf{c},h})=\underbrace{(w_{p_1}^{c_{1}(i_{1})}w_{p_2}^{c_{2}(i_{2})}\hdots
w_{p_l}^{c_{l}(i_{l})})w_{q}^{s_{0}^{\mathbf{v},\mathbf{r},d,h}},\hdots, (w_{p_1}^{c_{1}(i_{1})}w_{p_2}^{c_{2}(i_{2})}\hdots w_{p_l}^{c_{l}(i_{l})})w_{q}^{s_{2^{m}-1}^{\mathbf{v},\mathbf{r},d,h}}}\\
    \end{split}
 \end{equation}
 \end{strip}
 \!\!\!\!$\Psi(M^{\mathbf{v'},\mathbf{r},\mathbf{c},h})$ if  atleast one of $i_{k}\geq p_{k}$ where,  $0\leq i_{k}\leq2^{s_k}-1$ , $1\leq k\leq l$ and $0\leq j\leq 2^{m}-1$.  Therefore, after the trancation we left with a squence $\Psi_{Trun}(M^{\mathbf{v'},\mathbf{r},\mathbf{c},h})$  where, each elements of $\Psi_{Trun}(M^{\mathbf{v'},\mathbf{r},\mathbf{c},h})$ is of the form  $w_{q}^{g_{j}^{\mathbf{v'},\mathbf{r},h}}\!w_{p_1}^{c_{1}(i_{1})}\!w_{p_2}^{c_{2}(i_{2})}\!\hdots\!
w_{p_l}^{c_{l}(i_{l})}$\!\!  where, $0\!\leq i_{k}\leq\! p_{k}\!-\!1$ , 
 $0\!\leq\! j\!\leq\! 2^{m}\!-\!1$ and $1\leq\! k\!\leq\! l$.
   Clearly we can make $p_{1}p_{2}\hdots p_{k}2^{m}$ number of the elements of the form $w_{p_1}^{c_{1}(i_{1})}w_{p_2}^{c_{2}(i_{2})}\hdots
w_{p_l}^{c_{l}(i_{l})}w_{q}^{s_{j}^{\mathbf{v'},\mathbf{r},h}}$   if we vary all the $i_{k}$'s from $0$ to $p_{k}\!-\!1$ and $j$ from $0$ to $2^{m}\!-\!1$. Hence the length of  $\Psi_{Trun}(M^{\mathbf{v'},\mathbf{r},\mathbf{c},h})$ is 
 $p_{1}p_{2}\hdots p_{l}2^{m}$. We partition the length of $\Psi_{Trun}(M^{\mathbf{v'},\mathbf{r},\mathbf{c},h})$ by $\prod_{i=1}^{l}p_{i}$ parenthesis where, each parenthesis has  sequence of length $2^{m}$. Equation (8) represents the $i$-th parenthesis of $\Psi_{Trun}(M^{\mathbf{v'},\mathbf{r},\mathbf{c},h})$ and $\Psi_{Trun}(N^{\mathbf{v'},\mathbf{r},\mathbf{c},h})$ where, $i=i_{1}+\displaystyle\sum_{j=2}^{l}i_{j}\displaystyle\prod_{b=1}^{j-1}p_{b}$, $0\leq i_{j}\leq p_{j}-1$ and $1\leq j\leq l$. 
 \section{Proposed Construction of ZCCs}
\begin{theorem}
 Let $g:Z_{2}^m \rightarrow Z_{q}$ be a GBF as defined in \textit{Lemma} \textit{1}. Let $2\leq  p_i\leq 2^{s_i}$   and $\mathbf{c}=(c_1,c_2,\ldots,c_l)$ where $1\leq i\leq l$ and $0\leq c_{i}< p_{i}$. Then the set of codes\\
 $\bigg\{{\psi_{Trun}}(\Omega_{r}^{\mathbf{c}}),{\psi^{*}_{Trun}}(\Lambda_{r}^{\mathbf{c}}):0\leq r<2^{n},0\leq c_{i}\leq p_{i}-1\bigg\}$,
forms a $(\prod_{i=1}^{l}p_{i}2^{n+1},2^{m})-ZCCS_{2^{n+1}}^{2^m\prod_{i=1}^{l}p_{i}}$ if $h(\mathbf{v}')\!\in \{\lambda,\frac{q}{2}+\lambda\}$ $\forall \mathbf{v'}\in \{0,1\}^{n+1}$, where $\lambda\in Z_{q}$.
\end{theorem}
\begin{IEEEproof}
From (8) it can be observed that the $i$-th  parenthesis of $\Psi_{Trun}(M^{\mathbf{v'},\mathbf{r},\mathbf{c},h})$  can be expressed as  $\omega_{p_1}^{c_1 (i_1)}\omega_{p_2}^{c_2 (i_2)}\ldots\omega_{p_l}^{c_l (i_l)}\Psi(g^{\mathbf{v'},\mathbf{r},h})$   where, $i=i_{1}+\displaystyle\sum_{j=2}^{l}i_{j}\displaystyle\prod_{b=1}^{j-1}p_{b}$, $0\leq i_{j}\leq p_{j}-1$ and $1\leq j\leq l$. 
From (5), (8), \textit{Lemma} \textit{1} and \textit{Lemma} \textit{2} the ACCF between $\Psi_{Trun}(\Omega_{r}^{\mathbf{c}})$ and ${\Psi_{Trun}}(\Omega_{r'}^{\mathbf{c}'})$   for $\tau=0$ can be derived as
\begin{equation}
    \begin{split}
        &\Theta({\Psi_{Trun}}(\Omega_r^\mathbf{c}),{\Psi_{Trun}}(\Omega_{r'}^{\mathbf{c} '}))(0)\\
        &=\displaystyle\sum_{\mathbf{v'}}\Theta({\Psi_{Trun}}(M^{\mathbf{v'},\mathbf{r},\mathbf{c},h}),{\Psi_{Trun}}(M^{\mathbf{v'},\mathbf{r}',\mathbf{c}',h}))(0)\\
        &=\displaystyle\sum_{\mathbf{v'}}\Theta({\Psi}(g^{\mathbf{v'},\mathbf{r},h}),{\Psi}(g^{\mathbf{v'},\mathbf{r}',h}))(0)\prod_{d=1}^{l}\sum_{\alpha=0}^{p_{d}-1}\omega_{p_d}^{(c_d-c_d')\alpha}\\
        &=\Theta(\Psi(G_r),\Psi(G_{r'}))(0)\displaystyle\prod_{d=1}^{l}\displaystyle\sum_{\alpha=0}^{p_{d}-1}\omega_{p_d}^{(\mathbf{c}_d-c_d')\alpha}\\
        &=\begin{cases}
     p_{1}p_{2}\ldots p_{l}2^{m+n+1},& r=r',c=\mathbf{c}',\\
     0         ,& r=r', \mathbf{c}\neq\mathbf{c}',\\
     0         ,& r\neq r',\mathbf{c}=\mathbf{c}',\\
     0         ,& r\neq r',\mathbf{c}\neq \mathbf{c}'.
    \end{cases}
    \end{split}
\end{equation}
\newpage
Now, Using (5), (8), \textit{Lemma} \textit{1} and  the ACCF between ${\Psi_{Trun}}(\Omega_r^\mathbf{c})$ 
and ${\Psi_{Trun}}(\Omega_{r'}^{\mathbf{c}^{'}})$ for $0<|\tau|<2^m$ can be derived as,
\begin{equation}
    \begin{split}
        &\Theta({\Psi_{Trun}}(\Omega_r^\mathbf{c}),{\Psi_{Trun}}(\Omega_{r^{'}}^{\mathbf{c}^{'}}))(\tau)\\
        &=\Theta(\Psi(G_r),\Psi(G_{r^{'}}))(\tau)\prod_{d=1}^{l}\sum_{\alpha=0}^{p_d-1}\omega_{p_{d}}^{(c_d-c_d^{'})(\alpha)}\\
         &+\Theta(\Psi(G_r),\Psi(G_{r^{'}}))(\tau-2^m)\displaystyle\sum_{\alpha=0}^{p_{1}-2}\omega_{p_{1}}^{c_1(\alpha+1)-c_1^{'}\alpha}\\
        &\prod_{d=2}^{l}\displaystyle\sum_{\alpha=0}^{p_{d}-1}\omega_{p_{d}}^{c_d(\alpha)-c_d'(\alpha)}+\Theta(\psi(G_r),\psi(G_{r^{'}}))(\tau-2^m)\\
        &\displaystyle\sum_{f=1}^{l-2}\prod_{d=1}^{f}\omega_{p_{d}}^{c_{d}(0)-c^{'}_{d}(p_{d}-1)}\displaystyle\sum_{\alpha=0}^{p_{f+1}-2}\omega_{p_{f+1}}^{c_{f+1}(\alpha+1)-c^{'}_{f+1}(\alpha)}\prod_{k=f+2}^{l}\\
        &\displaystyle\sum_{\alpha=0}^{p_{k}-1}\omega_{p_{k}}^{c_{p_{k}}(\alpha)-c^{'}_{p_{k}}(\alpha)}+\Theta(\psi(G_r),\psi(G_{r^{'}}))(\tau-2^m)\\
        &\prod_{d=1}^{l-1}w_{p_d}^{c_d( 0)-c_d^{'}(p_d-1)}\displaystyle\sum_{\alpha=0}^{p_{l}-2}\omega_{p_{l}}^{c_{l}(\alpha+1)-c^{'}_{l}\alpha}.
    \end{split}
\end{equation}
 From \textit{Lemma} \textit{1}, we have, $\Theta(\Psi(G_r),\Psi(G_{r^{'}}))(\tau)=0$, $\forall\tau$, $0< |\tau|<2^{m}$. Therefore, from the above we can say,
\begin{equation}
 \begin{split}
    \Theta({\Psi_{Trun}}(\Omega_r^\mathbf{c}),{\Psi_{Trun}}(\Omega_{r^{'}}^{\mathbf{c} '}))(\tau)=0, 0<|\tau|<2^m.
 \end{split}
\end{equation}
From (9) and (11) We have,
\begin{equation}\label{thpoly9}
 \begin{split}
  \theta&({\Psi_{Trun}}(\Omega_r^\mathbf{c}),{\Psi_{Trun}}(\Omega_{r'}^{\mathbf{c} '}))(\tau)\\
  &=\begin{cases}
     p_{1} p_{2}\ldots p_{l}2^{m+n+1},& r=r',\mathbf{c}=\mathbf{c}',\tau=0,\\
     0         ,& r=r', \mathbf{c}\neq\mathbf{c}',0<|\tau|<2^m,\\
     0         ,& r\neq r',\mathbf{c}=\mathbf{c}',0<|\tau|<2^m,\\
     0         ,& r\neq r',\mathbf{c}\neq \mathbf{c}',0<|\tau|<2^m.
    \end{cases}
 \end{split}
\end{equation}
Similarly, it can be shown that
\begin{equation}\label{thpoly9}
 \begin{split}
  \theta&({\Psi^{*}_{Trun}}(\Lambda_r^\mathbf{c}),{\Psi_{Trun}^{*}}(\Lambda_{r'}^{\mathbf{c}'})(\tau)\\
  &=\begin{cases}
     p_{1} p_{2}\ldots p_{l}2^{m+n+1},& r=r',\mathbf{c}=\mathbf{c}',\tau=0,\\
     0         ,& r=r', \mathbf{c}\neq\mathbf{c}',0<|\tau|<2^m,\\
     0         ,& r\neq r',\mathbf{c}=\mathbf{c}',0<|\tau|<2^m,\\
     0         ,& r\neq r',\mathbf{c}\neq \mathbf{c}',0<|\tau|<2^m.
    \end{cases}
 \end{split}
\end{equation}
From \textit{Lemma} \textit{1}, we have 
$\Theta(\Psi(G_r),\Psi^*(\bar{G}_{r}))(\tau)=0,~ |\tau|<2^m.$
Therefore, from \textit{Lemma} \textit{1}, (5), (8) the ACCF between ${\Psi_{Trun}}(\Omega_r^\mathbf{c})$ and ${\Psi_{Trun}^{*}}(\Lambda_{r^{'}}^{\mathbf{c}^{'}})$ for $\tau=0$ can be derived as,
\newpage
\begin{equation}
    \begin{split}
    &\Theta({\Psi_{Trun}}(\Omega_r^\mathbf{c}),{\Psi_{Trun}^{*}}(\Lambda_{r^{'}}^{\mathbf{c}^{'}})(0)\\
    &=\displaystyle\sum_{\mathbf{v'}}\Theta({\Psi_{Trun}}(M^{\mathbf{v'},\mathbf{r},\mathbf{c},h}),{\Psi_{Trun}^{*}}(N^{\mathbf{v'},\mathbf{r}',\mathbf{c}',h}))(0)\\
&=\displaystyle\sum_{\mathbf{v'}}\Theta({\Psi}(g^{\mathbf{v'},\mathbf{r},h}),{\Psi^{*}}(s^{\mathbf{v'},\mathbf{r}',h}))(0)\prod_{d=1}^{l}\sum_{\alpha=0}^{p_{d}-1}\omega_{p_d}^{(c_d+c_d')\alpha}\\
&=\omega_{q}^{2\lambda}\Theta(\Psi(G_r),\Psi^{*}(\bar{G}_{r'}))(0)\displaystyle\prod_{d=1}^{l}\displaystyle\sum_{\alpha=0}^{p_{d}-1}\omega_{p_d}^{(c_d+c_d')\alpha}\\
&=0.
    \end{split}
\end{equation}
By the similar calculation as in  (10), we have
\begin{equation}
    \Theta({\Psi_{Trun}}(\Omega_r^\mathbf{c}),{\Psi_{Trun}^{*}}(\Lambda_{r^{'}}^{\mathbf{c}^{'}})(\tau)=0,\forall~ 0<|\tau|<2^{m}.
\end{equation}
Hence by (12), (13), (14) and (15) we  conclude the set
\begin{equation}
\begin{split}
    \bigg\{{\psi_{Trun}}(\Omega_{r}^{\mathbf{c}}),{\psi^{*}_{Trun}}(\Lambda_{r}^{\mathbf{c}}):0\leq r<2^{n},&0\leq c_{i}\leq p_{i}-1\bigg\},
\end{split}
\end{equation}
 forms a $(\prod_{i=1}^{l}p_{i}2^{n+1},2^{m})-ZCCS_{2^{n+1}}^{2^m\prod_{i=1}^{l}p_{i}}$.
\end{IEEEproof}
\begin{corollary}
Our construction gives optimal ZCCS of length of this form $(p_{1}p_{2}\hdots,p_{l})2^{m}$, where $p_{i}$'s are any prime number. From the fundamental theorem of arithmetic \cite{baker1984concise}   any number can be expressed as product of prime numbers therefore  the optimal ZCCS obtained by using our suggested construction gives all possible  length  of this form $n2^{m}$, where $n$ is any positive integer greater than or equal $1$. If $m=1$ we get all possible  even length optimal ZCCS.
\end{corollary}
\begin{remark}
 For $l=1$, the proposed result in Theorem 1 reduces to $(p2^{n+1}, 2^m)$-$ZCCS_{2^{n+1}}^{p2^m}$ as in [10]. Therefore, the proposed construction is a generalization of [10]
\end{remark}
\begin{corollary}(\cite{sarkar2020direct} )
 Let us assume that $G(h)$ is a path where, the
edges have the identical weight of $\frac{q}{2}$. Then $h(\mathbf{v'})$ can be expressed as
$$h(v_{0},v_{1},\hdots,v_{n})=\frac{q}{2}\sum_{\alpha=0}^{n-1}v_{\pi(\alpha)}v_{\pi(\alpha+1)}+\sum_{\alpha=0}^{n}u_{\alpha}v_{\alpha}+u,$$
where $u,u_{0},u_1,\hdots u_{n}\in \mathbb{Z}_{q}$. From (5), it is clear
that the $i$-th column of $\psi(\Omega_{r}^{\mathbf{c}})$ is obtained by fixing $\mathbf{{Y}}$ at  $\mathbf{{i}}\!\!\!=\!\!\!(i_0,i_1,\hdots,i_{m},\hdots,i_{{m+\sum_{i=1}^{l}s_i}-1})$, in the  expression of $M^{\mathbf{v'},\mathbf{r},\mathbf{c},h}$
where $(i_0,i_1,\hdots,i_{m},\hdots,i_{{m+\sum_{i=1}^{l}s_i}-1})$ is the binary representation of $i$.
Because the $i$-th column sequence of $\Psi(\Omega_r^\mathbf{c})$ is derived from a GBF whose graph is a path over n+1 vertices, hence from \cite{paterson2000generalized} the $i$-th column sequence of $\Psi(\Omega_r^\mathbf{c})$ is a q-ary Golay sequence. Thus each coloumn of $\Psi_{Trun}(\Omega_r^\mathbf{c})$ is Golay sequence. Thus the PMPER of each coloumn $\Psi_{Trun}(\Omega_r^\mathbf{c})$ is bounded by $2$. Similarly the PMPER of each coloumn of $\Psi^{*}_{Trun}(\Lambda_r^\mathbf{c})$ is bounded by $2$.
\end{corollary}
\begin{remark}
 From (6), it can be observed that $w_{q}^{M_{r'}^{\mathbf{v'},\mathbf{r},h}}$ is a root of the
polynomial: $z^{\sigma}-1$, where $\sigma$, denotes a positive
integer given by the least common multiple (lcm) of $p_{1},p_{2},\hdots,p_{l}$ and $q$. Therefore, the components of $\Psi(M^{\mathbf{v'},\mathbf{r},\mathbf{c},h})$ are given by the roots of the polynomial: $z^{\sigma}-1$.
\end{remark}
\begin{example}
Let us assume that $q\!=\!2$, $p_1\!=\!3$, $p_2\!=\!2$, $p_3\!=\!2$ $m=3$, $n=1$ and $s_1=2$, $s_2=1$ and $s_3=1$. Let us take the GBF $f:\{0,1\}^2\rightarrow\mathbb{Z}_2$ as follows:
 $f=y_1y_2+y_0$, where $G(f\arrowvert_{y_0=0})$ and $G(f\arrowvert_{y_0=1})$ give a path with $y_1$ as one of the end vertices. Let $h:\{0,1\}^{2}\rightarrow Z_{2}$ defined by $h(v_{0},v_{1})=v_{0}v_{1}$ From (4) we have,
 \begin{equation}
  \begin{split}
  & M^\mathbf{c}=y_1y_2+y_0+\frac{2c_1 }{3}(y_3+2y_4)+c_2 y_5+c_3 y_6,\\
  & N^\mathbf{c}=\bar{y}{_1}\bar{y}{_2}+\bar{y}{_0}+\frac{2c_1 }{3}(y_3+2y_4)+c_2 y_5+c_3 y_6,
  \end{split}
 \end{equation}
 where $c_1=0,1,2$, $c_2=0,1$ and $c_3=0,1$ From (5), we have
\begin{equation}
\begin{split}
 \Omega_r^\mathbf{c}&=\left\{M^\mathbf{c}+v_{0}v_{1}+v_0y_0+r_0y_0+v_{1}y_1:v_0,v_{1}\in\{0,1\}\right\},\\
 \Lambda_r^\mathbf{c}&=\left\{N^\mathbf{c}+v_{0}v_{1}+v_0\bar{y}_0+r_0\bar{y}_0+\bar{v}_{1}y_1:v_0,v_1\in\{0,1\}\right\},
 \end{split}
\end{equation}
where $(r_0)$ is the binary  representation of the integer $r$ and $0\leq r<2$. Therefore,
\begin{align*}
\begin{split}
    \Big\{{\Psi_{Trun}}(\Omega_{r}^{\mathbf{c}}),{\Psi^{*}_{Trun}}(\Lambda_{r}^{\mathbf{c}})&:0\leq r\leq 1, 0\leq c_1\leq 2,\\
    &0\leq c_2\leq 1,0\leq c_3\leq 1\Big\},
\end{split}
\end{align*} forms an optimal $(48,8)$-ZCCS$_{4}^{96}$ and the maximum coloumn sequence PMPER is at most 2.
\end{example}
\begin{remark}
 Our proposed construction also have some advantages over \cite{shen2021new}.
 \begin{enumerate}
     \item In \cite{shen2021new}  multivariable functions is used which is less feasible for hardware generation of sequences as their domains contain the domains of PBF as subset.
     \item Our construction  has more flexibility on the phases of sequences.
     \item Also in \cite{shen2021new} the length of ZCCs is of the form $q^{m}$ where $m\geq 2$, $q\in \mathbb{Z}^{+}$  which, may not produce all even lengths, for example 6.
 \end{enumerate}
 
\nocite{Nakamura}\nocite{deng_2000}\nocite{zcz_mimo}\nocite{zhangchao2006}\nocite{pke2015}\nocite{fan2008}\nocite{xli2009}\nocite{yli2014}\nocite{hmwang2007}
\end{remark}
\begin{table}[!t]
\centering
\caption{Comparison of the Proposed Construction with \cite{sarkar2018optimal,das2020new,sarkar2020direct,wu2020z,sarkar2021pseudo,
shen2021new} }\label{ointa}
\resizebox{\textwidth}{!}{
\begin{tabular}{|l|l|l|l|l|l|}
\hline
Source           & Based On       &Parameters  &Coditions &Optimal     \\ \hline
\cite{sarkar2018optimal} &Direct        &$(2^{k+p+1},2^{m-p})-ZCCS_{2^{k+1}}^{2^m}$     & $k+p\leq m$   &yes                             \\ \hline
\cite{das2020new}&Indirect         &$(K,M)-ZCCS_{M}^{K}$     &$K,M\geq 2$ &yes                                 \\ \hline
\cite{sarkar2020direct}  &Direct         &$(2^{n+p},2^{m-p})-ZCCS_{2^n}^{2^m}$   &$p\leq m$
& yes \\ \hline
\cite{wu2020z}   &Direct          &$(2^{k+v},2^{m-v})-ZCCS_{2^{k}}^{2^m}$     &$v\leq m,k\leq m-v$ &yes                                      \\ \hline
\cite{sarkar2021pseudo} &Direct         &$(p2^{k+1},2^m)-ZCCS_{2^{k+1}}^{p2^m}$     &$m\geq 2$, $k\leq m$, $p$ prime    &yes                                  \\ \hline
\cite{shen2021new} &Direct        &$(q^{v+1},q^{m-v})-ZCCS_{q}^{q^{m}}$ &$v\leq  m$         &yes                      \\ \hline
\cite{das2020new}&Indirect   &     $(K,M^{N+1})-ZCCS_{M}^{M^{N+1}P}$        &$K,M\geq 2$       &yes                           \\ \hline
\textit{Theorem 1} &Direct          &$(k2^{n+1},2^m)-ZCCS_{2^{n+1}}^{k2^m}$    &$ k,m,n\in \mathbb{Z}^{+}$            &yes                 \\ \hline
\end{tabular}}
\end{table}
\section{Conclusion}
In this work, we have proposed a direct construction of optimal ZCCSs for all possible even lengths using PBFs. The maximum column sequence PMEPR of the  proposed  ZCCSs is upper-bounded by $2$ which can be useful in MC-CDMA system to control high PMPER problem. The proposed construction also provides more flexible parameter as compared to the existing PBFs based constructions of optimal ZCCS.
\bibliographystyle{IEEEtran}
\bibliography{Bibliography.bib}
\end{document}